\documentclass{ab2018}
\usepackage{url}
\usepackage{graphicx}
\usepackage{natbib}

\newcommand{\OIII}{[O~{\sc iii}]}

\newcommand{\Ha}{H$_\alpha$}
\newcommand{\kms}{km\,s$^{-1}$}

\def\rmxaa{Rev. Mexicana Astron. Astrofis.}

\begin{document}

\title{Archaeology of Galactic Nuclei Activity\thanks{Based on the invite talk presented on the session of Physical Sciences Department Russian Academy of Sciences.}}

\titlerunning{Archaeology of Galactic Nuclei Activity}

\author{A.V. Moiseev, A. Arshinova, A.A. Smirnova }

\institute{Special Astrophysical Observatory, Russian Academy of Sciences, Nizhnij Arkhyz, 369167  Russia}

\authorrunning{Moiseev et al.}

\date{August 25, 2025/Accepted: September 1, 2026}
\offprints{Alexei Moiseev  \email{moisav@sao.ru} }

\abstract{
Considerable observational evidence suggests that the activity of supermassive black holes in galactic nuclei is transient. The term ``active galactic nuclei archaeology'' has even been coined. This implies the possibility of reconstructing the history of activity, such as changes in the nuclear  luminosity over time across various regions of the electromagnetic spectrum, by analysing how this activity manifested itself on galactic and extragalactic spatial scales. These phenomena include relic radio structures, gas clouds illuminated by the ``ionising echo'' of past activity, and Fermi/eROSITA bubbles. We provide a review of the results of galactic nucleus activity studies, focusing on its observable impact on the intergalactic medium and circumgalactic  environment. Our main focus is on optical observations of ionisation cones and evidence of switching between radiative (ionisation cones) and kinetic (radio jets) modes of nuclear activity.
\keywords{active galactic nuclei, galaxies, interstellar medium, high energy physics}
}

\maketitle

\section{Introduction}

Active galactic nuclei (AGN) are defined as ``nuclei whose  properties cannot be explained in the context  of normal star formation and evolution''~\citep{Terlevich1990RMxAA..21..163T}. The observed properties include both electromagnetic radiation and the acceleration of relativistic particles --- high-energy cosmic rays; moreover, in recent years, AGN have also been associated with neutrino fluxes \citep{Plavin2021ApJ...908..157P}. The AGN phenomenon has been known for more than 80~years, beginning with the pioneering work of Carl Seyfert, who discovered emission from highly excited gas in the nuclei of six nearby galaxies, while the broadening of the hydrogen Balmer lines corresponded to radial-velocity dispersions of 3500–8500\,\kms \citep{Seyfert1943ApJ....97...28S}. The subsequent discoveries in the 1960s of radio-loud star-like objects --- quasars \citep{Schmidt1963Natur.197.1040S}, radio galaxies, and the expansion of the list of Seyfert galaxies thanks to the Markarian survey of the sky~\citep{Markaryan1967Afz.....3...55M} triggered an  accumulation of new observational data and the development of theories about the nature of AGN. The bolometric luminosity of AGN can exceed $10^{47}$~erg\,s$^{-1}$ ($3\cdot10^{13}$ in solar luminosities), making them the most powerful sources of electromagnetic radiation in the observable Universe \citep{Trefoloni2023A&A...677A.111T}.

After many years of discussions, there is now no doubt that the phenomenon of activity is caused by the accretion of the surrounding matter onto the central supermassive black hole (SMBH). If only a few years ago the presence of an SMBH was implied  by sufficiently convincing, yet still indirect signatures  \citep[see the reviews by][]{Cherepashchuk2003,Cherepashchuk2016}, now radio 
interferometry has provided images of the SMBH in the nucleus of our own Galaxy \citep{EHT2022ApJ...930L..12E} and in the AGN M87 \citep{EHT2019ApJ...875L...4E}. A tight correlation has been found between the mass of the SMBH, measured by various methods listed in the reviews above, and the parameters of their host galaxies \citep[primarily the luminosity and mass of the spheroidal component, see ][] {Gebhardt2000ApJ...539L..13G,Ferrarese2000ApJ...539L...9F}, independent of the current level of nuclear activity. On the one hand, this allows one to consider the co-evolution of galaxies and SMBHs \citep{Kormendy2004ARA&A..42..603K}, and on the other hand, it indicates that SMBHs must be present in the centers of virtually all galaxies. For example, modern spectroscopic surveys such as DESI (Dark Energy Spectroscopic Instrument) make it possible to identify AGN and measure SMBH masses in large samples of dwarf galaxies \citep{Pucha2025ApJ...982...10P}. It is worth noting that the methods used here to measure the mass of the central object from a single spectrum go back to the pioneering works of E. A. Dibai in the 1970s–80s \citep{Dibai1977SvAL....3....1D}.

If the first estimates of the fraction of AGN showed only a few percent of Seyfert nuclei among galaxies in the nearby Universe, large spectroscopic surveys have shown that active nuclei are present in $\sim20\%$ of galaxies with masses above $10^{10}\,$M$_\odot$ \citep{Kauffmann2003MNRAS.346.1055K,Ho2008ARA&A..46..475H}. It follows that if every SMBH can become an AGN, then the total duration of the quiescent phase over a galaxy's lifetime turns out to be significantly longer than the active phase.

An obvious reason why galaxies cannot sustain nuclear activity continuously is that the central accretion engine exhausts its ``fuel'' rather quickly --- i.e., the gas contained within the region controlled by its gravity ($r\le 1$–10 pc). To maintain activity, even if gas is present in the galactic disk, it is necessary to transport it from orbits with radii of 1---20~kpc, which implies a roughly $10^4$ times decrease in angular momentum. Several dynamical processes capable of causing angular momentum loss in galactic gas are known: tidal interactions and galaxy mergers, central bars, circumnuclear spirals, binary supermassive black holes, etc. \citep{Combes2001sac..conf..223C}. Owing to the development over the last decade of optical 3D-spectroscopic sky surveys such as MaNGA (Mapping Nearby Galaxies at APO), a connection has been discovered between the AGN phenomenon and the presence of kinematically misaligned subsystems. In other words, the accretion of external gas by the galactic disk also facilitates accretion onto the SMBH \citep{Raimundo2023NatAs...7..463R,Zhou2025arXiv250700627Z}. This is also clearly observed in the most extreme case of misaligned subsystems — galaxies with polar rings, among which the fraction of AGN is enhanced \citep{SmirnovReshetnikov2020AstL...46..501S}; numerical simulations of gas transportation toward the nucleus in these systems have also been performed \citep{Khrapov2024OAst...3320231K}.

Unfortunately, in most cases it is difficult to determine unambiguously which angular-momentum transport mechanism operates in a given galaxy. Although numerous statistical estimates indicate some connection between the presence of an AGN and  factors such  as galaxy interactions, the presence of a strong bar, or a kinematically misaligned subsystem, this refers rather to a statistical trend than to a strict correlation. This can be explained by the fact that the timescale of the AGN duty cycle ($<~10^6$~years, see below) is significantly shorter than the characteristic timescale of large-scale dynamical processes ($10^8-10^9$ years).

Variations in the accretion rate, precession, and instability of the accretion disk lead to changes in the emission of the active nucleus across all spectral regions. Variability of AGN in the continuum and in optical emission lines was discovered as early as the beginning of the 1960s \citep{SharovEfremov1963IBVS...23....1S,Smith1963Natur.198..650S,DibaiPronik1967AZh....44..952D}. Starting with the pioneering work \citet{Cherepashchuk1973ApL....13..165C}, the time delay between variations in the continuum flux and in the broad emission lines has been used to measure the mass of the central object by the reverberation mapping method. 

Nuclear activity is considered one of the most important factors in models of galactic evolution due to the radiative and kinetic impact that AGN exert on the interstellar medium of the galaxy, up to the complete shutdown of star formation as a result of the loss of cold gas  \citep[``quenching'', see for review ][]{Silk2012RAA....12..917S,IllustrisTNG2018MNRAS.473.4077P,Silchenko2022}. For a proper understanding of the mutual influence between AGN and the host galaxy, as well as for constraining the parameters of numerical models of this process, it is necessary to estimate the duration of the full activity cycle and to study the traces of past activity in galaxies with quiescent nuclei. The first example here is our own stellar system --- the Milky Way. The relatively recent activity of the Galactic nucleus is linked to the presence of the hot-gas Fermi and eROSITA bubbles \citep{Predehl2020}, which are named after the respective space observatories operating in the gamma-ray and X-ray bands. Simulations show that both bubble systems, with diameters of 8 and 14 kpc, could originate from  a single outburst of nuclear activity about 2 Myr ago \citep{Yang2022}. This estimate is close to the age of the ionization cone directed perpendicular to the Galactic disk toward the Magellanic Stream --- a gaseous filament associated with the nearest massive satellites of our Galaxy \citep{Bland-Hawthorn2019,Fox2020ApJ...897...23F}. Interestingly, an analogous system of two symmetric hot bubbles about 7~kpc in size has been found with the same Fermi Gamma-ray Space Telescope in the nearest massive spiral galaxy, the Andromeda Galaxy \citep{Pshirkov2016MNRAS.459L..76P}.

In addition to the gamma-ray bubbles noted above, the known examples of nucleus activity turning on and off described in the literature can be divided into three classes of phenomena:

\begin{enumerate}
    \item Changing-look AGN (CL AGN), with characteristic activity durations of $\tau~\approx~1-10$ years.
    \item Relic radio structures, $\tau\approx10^6-10^8$ years.
    \item Relic ionized nebulae (ionization cones), $\tau\approx10^4-10^6$ years.
\end{enumerate}

In relation to the last two points, the term ``AGN archaeology'' has even appeared \citep{Morganti2017}. This implies the possibility of reconstructing the history of AGN radiation by analyzing its impact on galactic and extragalactic spatial scales. An obvious analogy arises with classical archaeology, in which the study of relic artifacts makes it possible to reconstruct  the greatness of bygone cultures.

Below, all the listed manifestations of AGN are briefly considered, with particular attention given to the recently discovered phenomenon of switching between different activity modes that are fundamentally distinguished by the accretion rates of matter onto the SMBH.

\begin{figure*}
\centering
\includegraphics[width=0.9\linewidth]{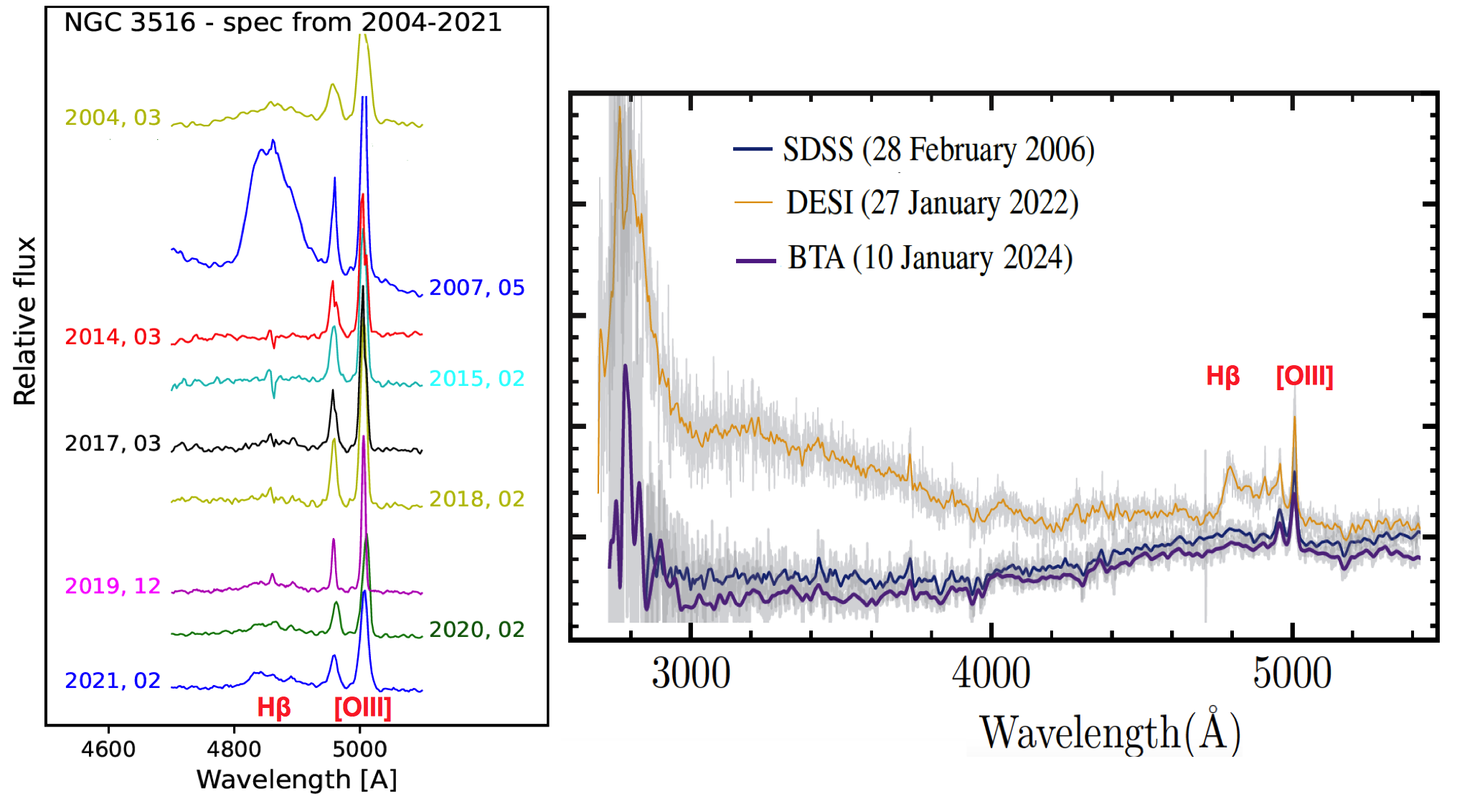}
\caption{Examples of optical spectra of changing-look AGN at different epochs; the positions of the broad H$_\beta$ emission line and the oxygen doublet \OIII$\lambda4959,5007$ are shown in red. {Left:} NGC~3516 in 2004--2021 \citep{Popovic2023A&A...675A.178P}, with most of the spectra obtained using the 1-m and 6-m telescopes of SAO RAS.  Right:  the galaxy at redshift $z=0.34$, J075947.73+112507.3.  {Shown are} the spectra from the SDSS~(2006), from DESI~(2022), and from the 6-m telescope~(2024) according to \citet{Guo2025A&A...698A.135G}. Credit: Astronomy \& Astrophysics.}
\label{fig:Moiseev_1}
\end{figure*}

\section{Changing-look AGN}

In the case of changing-look AGN, spectra obtained at different epochs show either a complete disappearance of the signatures of activity in an object previously classified as a classical quasar or Seyfert nucleus, or, conversely, the sudden appearance of characteristic broad lines in the spectrum of a previously ``quiescent'' nucleus, indicating the presence of an AGN \citep{Ruan2016}. Alternatively, the phenomenon may involve switching between the Seyfert-1 and Seyfert-2 activity types --- i.e., the disappearance or appearance of broad hydrogen Balmer lines in the presence of bright forbidden emission lines of other elements. Interestingly, among the six galaxies included in the list of C.K. Seyfert, two belong to the CL AGN type: NGC~3516 \citep[fig.\ref{fig:Moiseev_1},][]{Popovic2023A&A...675A.178P} and NGC~4151 \citep{Feng2024ApJ...976..176F}. Such a high number of detections, significantly larger than in other samples, may be explained by the fact that these classical objects have the longest spectral time series available, spanning several decades. Among other nearby galaxies of this type, NGC~1566 has been studied in great detail~\citep{Oknyansky2019MNRAS.483..558O}.

Changes in the type of activity on a timescale of several years are not consistent with the classical unified model of an active nucleus, since the characteristic viscous timescale of variations in the accretion disk is estimated to be $10^3-10^4$ years. However, several authors have suggested that the observed changes are associated with a much shorter thermal timescale \citep{Graham2020MNRAS.491.4925G}.
Various models have been proposed to explain the changing-look behavior of AGN, such as the redistribution of absorbing matter in the central region, including a dusty nuclear wind, accretion-disk instabilities, and genuine changes in the accretion rate. Yet even in the latter case it is not currently possible to assert whether the phenomenon represents a global shutdown of activity for hundreds of thousands or a million years (as in our Milky Way), or whether a reverse transition will occur soon. It is also possible that some fraction of changing-look AGN cases are related to tidal disruption events (TDEs), which have been actively studied in recent years, partly owing  to the mission of the X-ray space observatory Spektr-RG \citep{Sazonov2021}. These events are thought to be associated with individual stars that have come too close to the SMBH, causing  a short-lived accretion disk to form from the resulting tidal debris of hot plasma.

Hopes for further progress in understanding the nature of CL AGN are related to large spectroscopic sky surveys that add new observational epochs to the existing SDSS data. First of all, this concerns the DESI survey mentioned above, during which very interesting objects have already been discovered. For example, the DESI spectrum obtained in 2022 for the galaxy J075947.73+112507.3 suggests  the presence of an active nucleus, whereas two years later, as demonstrated by the spectrum from the 6-m telescope of SAO RAS (fig.~\ref{fig:Moiseev_1}), the broad lines characteristic of AGN have completely disappeared, just as they had earlier in 2006 during the SDSS epoch. At the same time, the nuclear light curve over this period shows between 4 and 6 flares. The nature of such rapid changes in the accretion rate remains unclear~\citep{Guo2025A&A...698A.135G}.

\section{Relic radio structures}

A significantly longer characteristic timescale of active events is provided by observations in the radio band. In a number of radio-loud AGN, it is possible to trace extended relic structures that indicate past activity. They are distinguished by the steep slope of the synchrotron radio spectrum, from which, under certain assumptions about radiative losses, one can estimate the age of the last injection of relativistic electrons from the radio jet into the magnetic field, as well as by specific features in the spatial distribution of the radio emission (the presence of a pronounced compact core, etc.). Here we are dealing with timescales from 1 to 100 Myr, consistent with the spatial sizes of the observed structures.
A detailed review of  the results in this area is presented in \citet{Morganti2017} (where the term ``archaeology of active galaxies'' was introduced) and \citet{Morganti2024}. Substantial progress in such observations is associated with the commissioning of low-frequency, highly sensitive radio telescopes with high angular resolution, such as  the Low-Frequency Array (LOFAR). Observing frequencies below 250 MHz are important for detecting the oldest structures in terms of synchrotron-lifetime considerations, since their emission intensity depends on frequency as $I_\nu\sim\nu^{-\alpha}$, $\alpha~>~1$.
The occurrence rate of restarted activity in the radio domain (restarted AGN) in deep observations with LOFAR is about 13–15\% \citep{Jurlin2020A&A...638A..34J}. Such observations make it possible to trace an entire sequence of radio-jet activity episodes. For example, in J1225+4011 one observes traces of three pairs of radio structures with sizes of 118, 572, and 1349 kpc, corresponding to kinematic ages of 2, 19, and 220 Myr, respectively \citep{Chavan2023MNRAS.525L..87C}.

\section{Ionization echo}

The unified model of an active nucleus assumes that the accretion disk around the SMBH is surrounded by an opaque gas–dust torus. In the case of the nearest AGN, these tori, whose sizes range from 0.5 to 10~pc, can be directly observed using interferometry in the infrared at wavelengths of $8-13\,\mu m$ \citep{Vermot2021A&A...652A..65V}. Collimation by the torus of the radiation from the hot accretion disk leads to the ionizing UV radiation propagating in the form of two symmetric cones. If gas clouds fall within the cone, whether in the galactic disk or in its surroundings, then the cone -- or a portion of it -- can be directly observed in optical emission lines, the brightest of which is typically the \OIII$\lambda4959,5007$\AA\ doublet. A review of the current state of studies of extended (10~--~100~kpc) ionization cones is presented in our recent work~\citet{MoiseevSmirnova2023}.

\begin{figure*}
\centering
\includegraphics[width=0.9\linewidth]{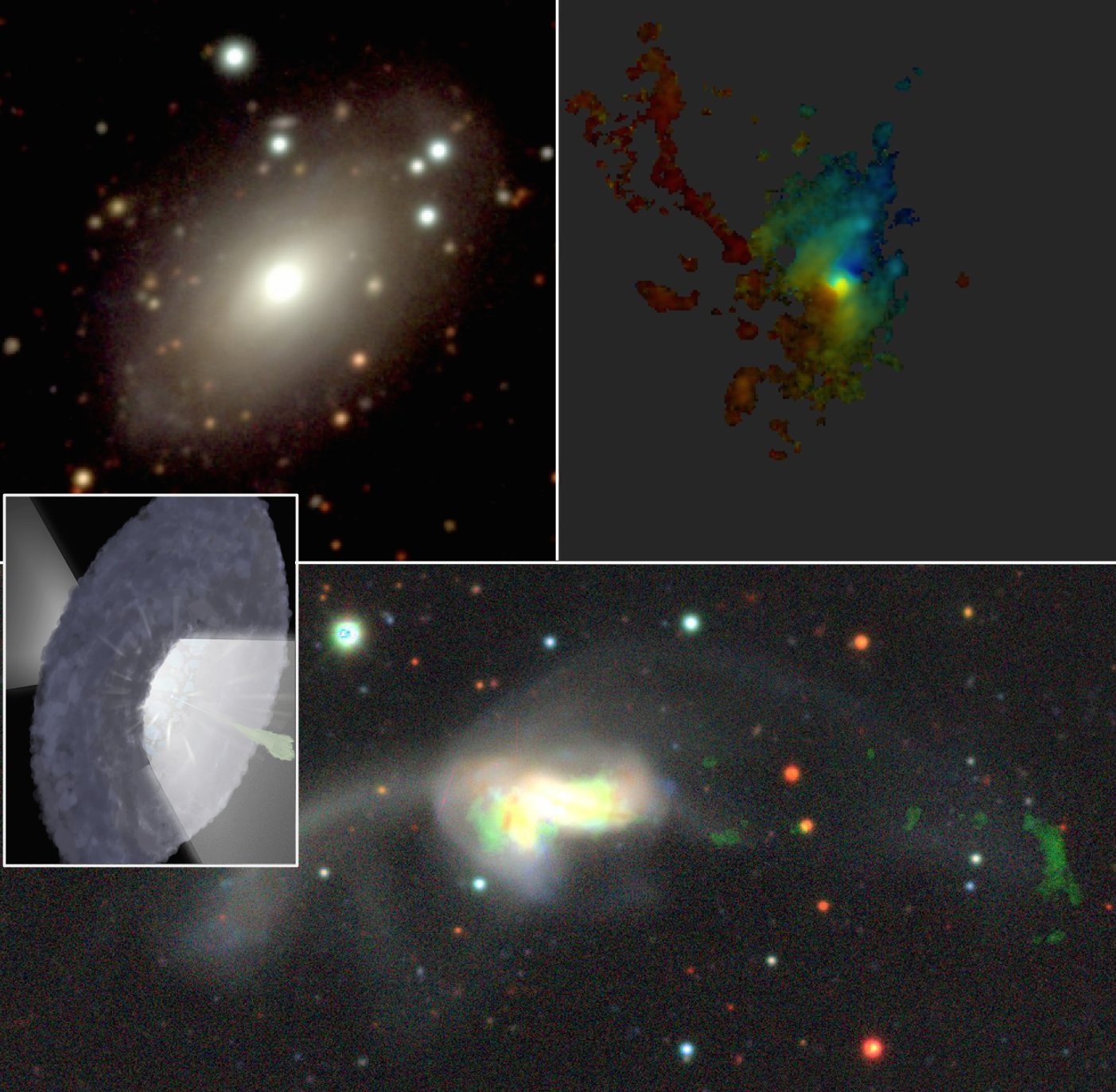}
\caption{Examples of ionization cones. Top: a deep optical image of Mrk~6 obtained with the 6-m telescope of SAO RAS (left) and the distribution of ionized gas on the same scale (right), according to \citet{Smirnova2018MNRAS.481.4542S}. The brightness corresponds to the logarithm of the \Ha\ line intensity, while the color represents the gas radial velocities ($-100...+200$\,\kms for the blue and red, respectively). Bottom: an image of NGC~5514 from the DESI Legacy Survey. The distribution of emission in the ionized oxygen line is shown in green, based on observations with the MaNGaL instrument mounted on the 2.5-m telescope of SAI MSU. The inset shows an artistic illustration of an ionization cone, artwork by E.A.~Malygin (SAO RAS).}
\label{fig:Moiseev_2}
\end{figure*}

Fig.\ref{fig:Moiseev_2} shows one of the typical examples of ionization cones --- the case of the nearby galaxy Mrk~6, where filaments of ionized gas have been detected using the 6-m telescope of SAO RAS, extending out to distances of 40~kpc from the nucleus, beyond the stellar disk. The observed distribution of radial velocities can be explained under the assumption that the AGN radiation has ionized the gas that was captured by Mrk~6 from the intergalactic medium \citep{Smirnova2018MNRAS.481.4542S}.

If the production of ionizing photons by the nucleus decreases significantly, an ionization echo effect is observed: outer gas clouds exhibit AGN-like signatures, while the nucleus itself no longer shows them. If the recombination timescale in the lines is sufficiently short, then the characteristic fading time equals the light-travel time from the nucleus to the ionized cloud. The prototype of this class of objects is Hanny's Voorwerp --- a nebula discovered in SDSS images by the volunteer of the Galaxy Zoo project, Hanny van Arkel, next to the spiral galaxy IC~2497. The spectrum of the nebula contains characteristic AGN lines, while the infrared luminosity indicates that the nucleus itself shows no activity and is not simply obscured by dust. The observed picture is explained by assuming that the activity of the ionizing source has decreased by almost two orders of magnitude over the past $10^5$ years \citep{Lintott2009}.

Initially, detection  of ionization echoes was occasionally, but in order to draw reliable conclusions about the history of the ionizing activity of AGN, one must work with statistically significant samples of objects. We will focus on the most important results in this field, referring the reader to the review~\citet{MoiseevSmirnova2023} for further details.

Among the 20 galaxies with extended emission structures discovered by volunteers of the Galaxy Zoo project and confirmed by spectroscopic observations, 7 show evidence of fading activity  including Hanny’s Voorwerp  \citep{Keel2012}. Further investigation of their activity histories was carried out using data from the Hubble Space Telescope  \citep[emission-line imaging and spectroscopy ---][]{Keel2015,Keel2017,Harvey2023}, by means of 3D spectroscopy with the 6-m telescope of SAO RAS \citep{Keel2015}, with the 8-m telescopes of the Gemini Observatory and the VLT \citep{Keel2017,Finlez2025arXiv250701115F}, as well as using archival data from the MaNGA survey \citep{Mo2024MNRAS.529.4500M}. Spectroscopic observations are necessary to demonstrate that the observed gas is dynamically cold and is ionized specifically by radiation from the nucleus. It has been shown that in these galaxies the nuclear luminosity has varied by a factor of 0.2--3 dex over the past $(4-8)\cdot10^4$ years \citep{Finlez2025arXiv250701115F}.

\begin{figure*}
\centering
\includegraphics[width=0.9\linewidth]{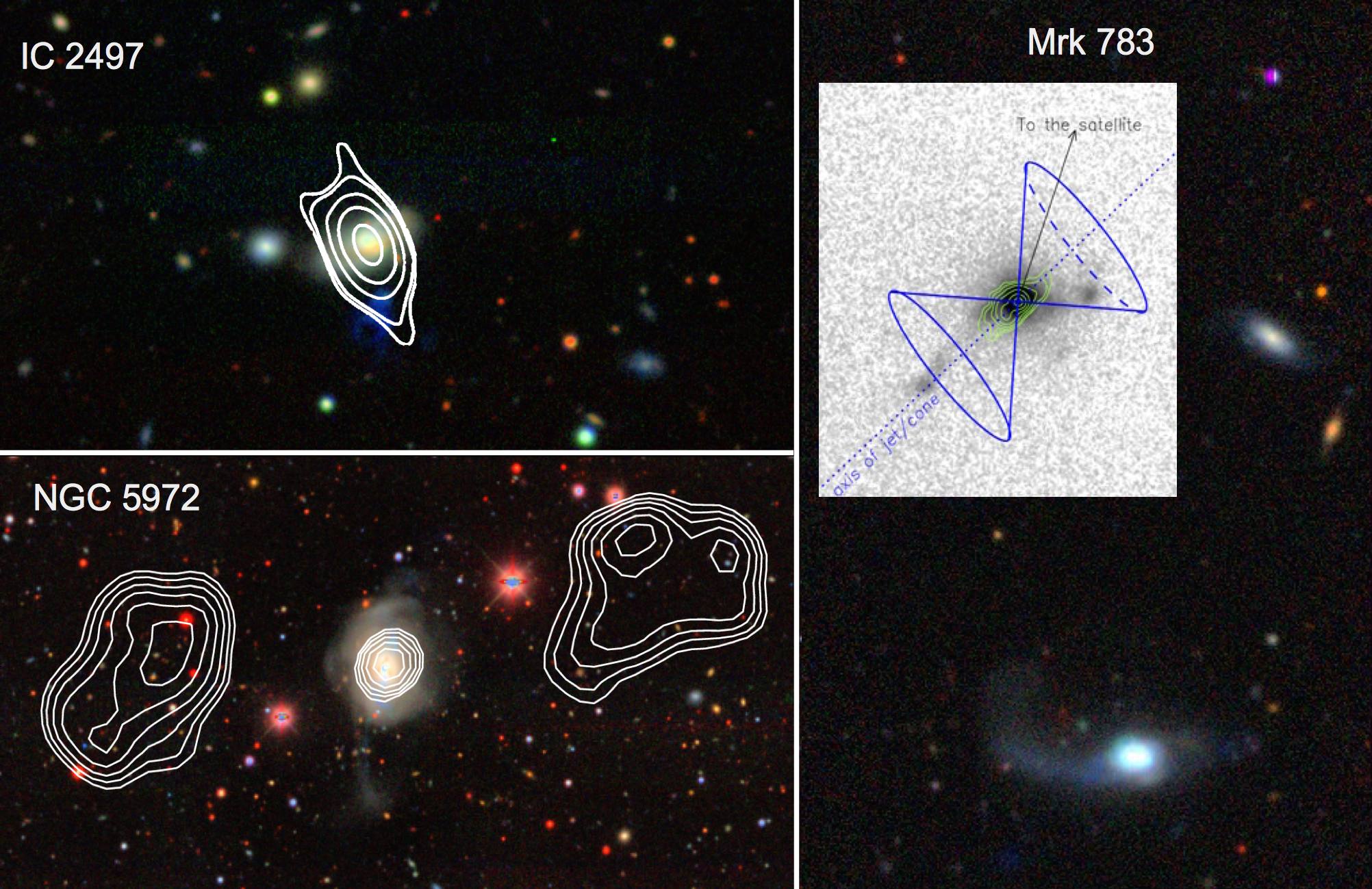}
\caption{Optical images of galaxies exhibiting accretion-mode switching from the DESI Legacy Survey. For IC~2497 and NGC~5972, contours of synchrotron radio emission from \citet{Jozsa2009} and from the archive of the NRAO/VLA Sky Survey in the L band, respectively, are shown. In the inset for Mrk~783, an \OIII\ image obtained with the 2.5-m telescope of SAI MSU is presented, with overlaid radio contours and a schematic illustration of the ionization-cone geometry \citep{Moiseev2023Univ....9..493M}.}
\label{fig:Moiseev_3}
\end{figure*}

A targeted search for external clouds ionized by nuclear radiation among nearby galaxies ($z < 0.03$), at a surface brightness level an order of magnitude lower than that of Hanny’s Voorwerp, within the framework of the TELPERION project (Tracing Emission Lines Probing Extended Regions Ionized by Once-active Nuclei), has led to the discovery of three additional cases of fading activity in a luminosity-limited sample of 111 AGN \citep{Keel2022MNRAS.510.4608K}. The most striking example is the interacting system NGC~5514, in which the ionizing activity has decreased by more than a factor of 3 over the past 250 thousand years, shown in fig.~\ref{fig:Moiseev_2}. It has been possible to estimate the occurrence rate of such clouds: they are detected in $\sim 2\%$ of nearby AGN, and this fraction increases to 12\% among all active galaxies that display prominent tidal structures, since in these systems the probability of having a suitable ``screen'' tracing the ionization cone is higher \citep{Keel2024}.

Among 93 AGN that have undergone a recent starburst in the SDSS/MaNGA survey, six cases of extended emission regions have been identified, five of which show evidence of declining ionizing activity. In all cases the relevant timescales are shorter than 100 thousand years, which is a consequence of the limited field of view of the spectrograph~\citep{French2023ApJ...950..153F}.
It should be noted that the potential of the MaNGA survey, which has obtained 3D spectroscopic data for more than 10\,000 galaxies, has not yet been fully realized in this direction.

Although only about a dozen galaxies with identified signatures of declining ionizing activity are known at present, this list will continue to grow thanks to the new deep sky surveys. For example, in the first publicly released images of the LSST, five new extended ionization cones have already been discovered~\citep{Jonkeren2025RNAAS...9..201J}. The main difficulty remains the need for follow-up spectroscopy of low-surface-brightness objects, which requires telescopes with diameters of no less than 3–4 meters.

\section{Galaxies with activity-mode switching}

In the literature, signatures of fading are usually considered separately for radio-loud and radio-quiet AGN, since these types of objects are associated with different accretion rate (low or high rate, the latter being close to the Eddington limit), or with kinetic and radiative modes of activity, respectively \citep{Morganti2017}. However, thanks to the accumulation of multiwavelength data for nearby AGN, it has been found that a galaxy nucleus may switch from a kinetic to a radiative mode, or vice versa. A detailed theoretical explanation of such behavior does not yet exist, but it is likely that an important role is played by the impact the nucleus exerted on the interstellar medium of the galaxy during the preceding phase of activity \citep{HopkinsElvis2010}. So far, this concerns only a few galaxies, discussed below. Optical and radio images of three of them are shown in fig.~\ref{fig:Moiseev_3}.

Thus, in Hanny’s Voorwerp --- IC~2497, traces of a relic radio jet about 100 million years old have been detected \citep{Smith2022}. It is possible that the galaxy Mrk~783, recently studied by us with the 2.5-m telescope of SAI MSU and the 6-m telescope of SAO RAS, represents an example of a transition toward the state observed in Hanny’s Voorwerp \citep{Moiseev2023Univ....9..493M}. This system contains an extended old radio structure, an ionization cone at least $10^5$ years old that affects the disk of a companion galaxy at a projected distance of about 100~kpc, and an inner compact radio jet whose direction does not coincide with either of these structures.

In NGC~5972 a fading ionization cone (on a timescale of several tens of thousands of years) coexists with a perpendicular relic structure typical of a radio galaxy whose activity ceased no less than several tens of millions of years ago \citep{Harvey2023}. Such a misalignment of spatial orientations may be caused by the presence of a binary SMBH in the nucleus.
In the radio galaxy Mrk~1498, a relic radio structure (with an age of about 100 million years), ongoing radio activity, and a fading ionization cone not aligned with the old radio structure likewise coexist \citep{Cazzolli2024}.

An interesting example is the more distant object RGZ~J123300.2+060325 (redshift $z~=~0.3$), in which an extended radio jet about 6 million years old is observed simultaneously with an ionization cone that indicates a strong decrease in the UV luminosity of the nucleus over the past 150 thousand years \citep{Sanderson2024}. This galaxy belongs to the type known as a ``green bean''. These objects are considered related to Hanny’s Voorwerp, but exhibit an order of magnitude higher luminosity, representing ionization echoes of powerful quasars rather than Seyfert galaxies. In the SDSS survey only 17 such ``beans'' were found, located at redshifts $z = 0.2-0.6$ \citep{Schirmer2013}. In optical images they appear elongated and bright green due to the significant redshift of the \OIII{} line into the $r$-band. This list was recently expanded to 166 ``green bean'' candidates, 5 of which have already received spectroscopic confirmation as genuine ionization cones \citep{Dijeau2025AJ....169..207D}.

For the purpose of obtaining new observational data on objects caught at the rare moment of switching their activity mode, in 2025 we started a program on the SAO RAS 6-m telescope  aimed at a comprehensive study of the kinematics, morphology, and ionization state in a sample of ``green bean'' galaxies with extended radio structures  \citep{Arshinova2026CoSka..56a..31A}.
 
\section{Conclusion}

In the context of the ``archaeological study'' of galactic activity, we highlight the following points as the most important in our opinion:

\begin{enumerate}
    \item Observations of extended (and distant) emission clouds make it possible to obtain information on the kinematics and ionization state of gas at distances up to 100~kpc from active galaxies, and to study the ionizing activity of their nuclei on timescales of $10^4-10^6$ years.
    \item In a number of nearby ($z<0.3$) active galaxies we see evidence of transitions between different accretion modes (dominance of a radiation cone or of a radio jet) on a characteristic timescale of $\sim10^5$ years. A change in the accretion rate may be related to the impact exerted on the interstellar medium of the galaxy during the preceding stage of activity.
\end{enumerate}

\section*{Fundings}
This work was carried out as part of the SAO RAS government contract approved by the Ministry of Science and Higher Education of the Russian Federation.  Part of the results presented here were obtained from observations with the 6-m telescope of SAO RAS, conducted with the support of the Ministry of Science and Higher Education of the Russian Federation. The renovation of telescope equipment is currently provided within the national project ``Science and Universities''.

\bibliographystyle{aa}

\end{document}